\documentclass[12pt]{spieman}  
\usepackage{amsmath,amsfonts,amssymb}
\usepackage{graphicx}
\usepackage{setspace}
\usepackage{tocloft}

\title{A Study on Smart Online Frame Forging Attacks against Video Surveillance System}

\author[a]{Deeraj Nagothu}
\author[a]{Jacob Schwell}
\author[a,*]{Yu Chen}
\author[b]{Erik Blasch}
\author[c]{Sencun Zhu}
\affil[a]{Dept. of Electrical \& Computer Engineering, Binghamton University, Binghamton, NY 13902}
\affil[b]{The United State Air Force Research Laboratory (AFRL), Rome, NY 13441, USA, 13441}
\affil[c]{Dept. of Computer Science and Engineering, Penn State University, University Park, PA 16802}

\cftpagenumbersoff{figure}
\cftpagenumbersoff{table} 

\begin{document} 
\maketitle

\begin{abstract}
Video Surveillance Systems (VSS) have become an essential infrastructural element of smart cities by increasing public safety and countering criminal activities. A VSS is normally deployed in a secure network to prevent the access from unauthorized personnel. Compared to traditional systems that continuously record video regardless of the actions in the frame, a smart VSS has the capability of capturing video data upon motion detection or object detection, and then extracts essential information and send to users. This increasing design complexity of the surveillance system, however, also introduces new security vulnerabilities. In this work, a smart, real-time frame duplication attack is investigated. 
We show the feasibility of forging the video streams in real-time as the camera's surroundings change. The generated frames are compared constantly and instantly to identify changes in the pixel values that could represent motion detection or changes in light intensities outdoors. An attacker (intruder) can remotely trigger the replay of some previously duplicated video streams manually or automatically, via a special quick response (QR) code or when the face of an intruder appear in the camera field of view.  
A detection technique is proposed by leveraging the real-time electrical network frequency (ENF) reference database to match with the power grid frequency.

\end{abstract}

\keywords{Video Surveillance, Frame Duplication Attacks, Edge Computing, Electrical Network Frequency (ENF)}

{\noindent \footnotesize\textbf{*}Corresponding Author: Yu Chen,  \linkable{ychen@binghamton.edu} }

\section{Introduction}
\label{sec:intro}  

Video Surveillance Systems (VSS) have become an essential infrastructural element of smart cities by increasing public safety and countering criminal activities \cite{collins2000system, nikouei2018real, yu2018pulic}. Surveillance cameras are deployed in large numbers to support modern infrastructures and ever-expanding smart cities and communities \cite{chen2016smart, nikouei2018smart}. Considering the increasing importance of VSS in public safety applications, the number of attacks on the VSS has grown quickly in recent years \cite{costin2016security}. ``Hiding in the plain sight'' has become the new form for visual layer attacks. While VSS has proven to be an effective security system to deter crimes and provide evidences for forensics analysis, the more advanced and populated a security feature is, sometimes the more susceptible it is to the attacks \cite{mendez2017internet, nagothu2018microservice}.

The evolution of surveillance camera networks started from transferring live video feed through direct serial connection to a network-based deployment \cite{xu2018real, nikouei2018eiqis}. The network communication also makes it vulnerable to network-based attacks like Man-in-the-Middle or Distributed Denial of Service (DDoS) attacks \cite{obermaier2016analyzing}. Gaining unauthorized access over the surveillance network could equally prove to be a significant security breach or in some cases more dangerous. Many remote sites solely rely on live video feed for the security of the premises, such as power generating stations, water dams, or locations that are not feasible for human operators to access regularly  \cite{chen2018smart, wu2015pseudo}. 

Frame duplication attacks are one of the most common attacks since they do not involve disturbing the state of the surveillance system and cause less suspicion \cite{singh2015detection, ulutas2017frame}. Many video altering software techniques have made it difficult to distinguish between forged and real video stream. The attacks discussed so far involved using a pre-recorded video stream to mask the current events and making the surveillance zone vulnerable \cite{bouchrika2018survey, matusek2008nivss}. In any case, a pre-recorded video could still be detectable by human perception if it does not match the constant environmental changes. 

The modern Internet of things (IoT) is more computationally capable than the techniques of a decade ago, and a compromised network of IoT could provide an attacker with enough on-board computation support to modify the video stream without using a rogue server for master-slave control. In this work, we study the implementation of such attacks performed on networked surveillance cameras where the frame duplication attack is performed using a real-time pre-recorded stream. The pre-recorded stream is modified promptly based on changes in environment like light intensity or object displacement. 

Our contributions are presented as follows.

\vspace{-10pt}
\begin{itemize}
    \item Developing a real-time smart attack model to demonstrate the feasibility of visual-data layer attack at the edge;
    \item Implementing and testing a remote or automatic attack triggering mechanism based on face detection or manual operation; and 
    \item Proposing an online detection system using electrical network frequency (ENF).
\end{itemize}
\vspace{-10pt}

The rest of the paper is organized as follows. Section \ref{sec:background} is a brief survey of the literature. Section \ref{sec:theory} introduces the algorithm control flow and stages of attack implementation. Section \ref{sec:testbed} presents the test-bed setup. Section \ref{sec:enf} proposes a detection technique using ENF as a reference signal. Section \ref{sec:conclusion} concludes with future work and discussions.

\section{Background Knowledge and Related Work}
\label{sec:background}
Frame duplication attacks against video surveillance systems have been a threat for a long time, and researchers have developed various spatial and temporal analysis frameworks to detect the duplicated frames \cite{wang2007exposing}. These algorithms mostly extract features from a video sub-sequence and compare them with other sub-sequences for similarity \cite{singh2015detection}. 
A number of correlation techniques \cite{ulutas2017frame, wahab2014passive, fadl2018authentication} have also been adopted to identify frame duplication and region duplication in a video. All these similarity detection techniques look up in the stored surveillance recording database, and hence they require much computation time to process each video frame. 
This is especially true when the duplicated video source changes regularly. Moreover, for surveillance cameras with higher recording resolution, it could cost even more processing time to achieve a higher detection rate. 

Even though some algorithms can extract relevant features and make a thorough comparison of video sub-sequences, the analysis are often conducted offline to serve the purpose of forensics. For the lack of a real-time frame duplication detection method, the attacker is able to accomplish her goal as long as the false frames are capable to deceive human perception. In addition, the success of a frame duplication attack normally means the missing of the genuine footage, which makes even the forensics analysis a challenging job.


Many outdated camera firmware can be exploited via various network attacks, resulting in software backdoor in camera firmwares \cite{heffner2013exploiting, shoshitaishvili2015firmalice}. A zero-day exploit could lay dormant in the firmware, allowing the attacker to access the camera and manipulate the video stream whenever required \cite{seralathan2018iot, xie2017vulnerability}. While this visual-data layer attack can be triggered through any mechanism \cite{costin2016security}, here we consider two different types: \textit{manual trigger} and \textit{visual data trigger}. Manually recording and replaying a video can be performed with more control, but it would require a persistent connection to a remote command and control server. While it is possible to detect unsolicited network traffic to an unknown server using appropriate firewall rules, a stealthy attack, which is triggered without any constant communication to a remote server, could remain undetected. Differently, a visual data trigger, as the attack launch command, can be any object in the camera view, e.g., intruder's face or a special QR code. The attack code may either mask live feed using pre-recorded videos or pixelate the targeted user (e.g., an intruder) to keep him/her unrecognized, depending on the application scenario.

In the following section, we will discuss the implementation of the frame duplication attack on both video and audio feeds at the edge.

\section{Real-Time Frame Duplication Attack at the Edge}
\label{sec:theory}

\subsection{System Overview}
An automated visual layer attack happens when an unauthorized person (also called intruder or attacker) shows up in the surveillance vicinity. We assume that the edge device responsible for surveillance (i.e. camera) has been compromised by the attacker through network attacks. The attacker can not only access the real-time surveillance feed, but also cut in by replaying some pre-generated audio-video stream to hide his appearance in the surveillance camera. The attack is divided into two parallel threads, one for determining the attack trigger point and the other for launching the video-audio replay attack. 

 \begin{figure} [ht]
   \begin{center}
   \begin{tabular}{c} 
   \includegraphics[height=13cm]{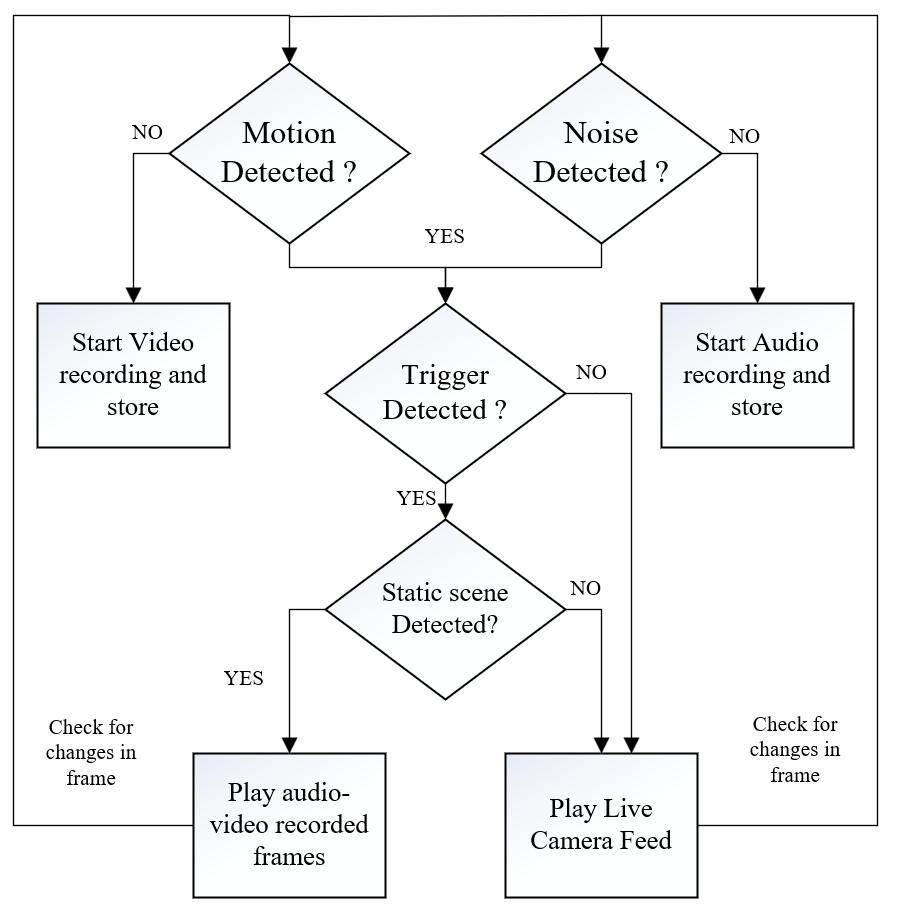}
	\end{tabular}
	\end{center}
   \caption[example] 
   { \label{fig:state_diagram} 
Flow Diagram of the Frame Duplication Algorithm.}
   \end{figure} 
   
Figure \ref{fig:state_diagram} shows the flow diagram of our algorithm. The conditions in the diagram determine whether there is a trigger in the camera, if so when to launch a replay attack, and whether to store the current stream into the database for potential future replay.  Upon detecting a trigger, the \textit{video replay module} initializes the frame duplication attack by using the most recent recording. If there is no motion detected in the current scenario, then the algorithm starts a new recording. 
The \textit{motion detection module} first looks for motions in the camera view by performing Gaussian Blur on the frames to smooth out edges and minimize errors due to noise. 
The \textit{audio replay module} runs in parallel to the video replay module to detect noise in the environment and record a static audio with no background noise. When the replay attack is triggered, both video frames and the audio samples are combined as a single synchronised media recording and used to mask the currently ongoing events.

Next we describe our attack algorithms in details.

\subsection{Video-Audio Replay Attack}

Recent studies have indicated that detecting the frame duplication attack in a static scene is difficult by comparing the objects in replayed motion \cite{fadl2018authentication}. The difficulty is partly due to the noise interference of the camera over its video output, as well as the changes in the light intensity caused by indoor lighting or natural ambient light. To exploit the limitations of existing detection techniques and human perception of live video streams, in our attack, the compromised surveillance camera retains a recently recorded static scene for masking live stream whenever triggered. The recorded static scene is updated continuously to keep up with the changes in the environment, like objects displaced or varying light intensities. 

The motion detection algorithm detects the number of changes in the intensities of pixels in subsequent frames. For different sensitivities of the environment, a different threshold, i.e. number of pixel values changed, can be used to detect motions. For outdoor-based motion detection, one may use a higher threshold due to changes in naturally occurring events or a distant moving object, whereas for indoor-based applications, the threshold can be smaller because the scenes are mostly static, except the cases of ambient light changing or people passing by. To reduce the computational time and motion detection sensitivity, a Gaussian blur is performed on incoming frames to smooth out edges and minimize errors due to noise. The Gaussian blur performs convolution on the image, acting as a low pass filter and therefore attenuating high-frequency components more than the lower-frequency components. Since human movement in the camera view appears as a low-frequency change while noise is a high-frequency change, removing the noise helps the algorithm better distinguish human motions from noise.

\[ G(x,y) = \frac{1}{2 \pi \sigma^2} e^{- \frac{x^2 + y^2}{2 \sigma^2}}  \]

\noindent where \( \sigma^2 \) is the variance of the Gaussian distribution, and x and y are the distances from the origin in the horizontal axis and the vertical axis, respectively.

\begin{figure} [ht]
   \begin{center}
   \begin{tabular}{c} 
   \includegraphics[height=8cm]{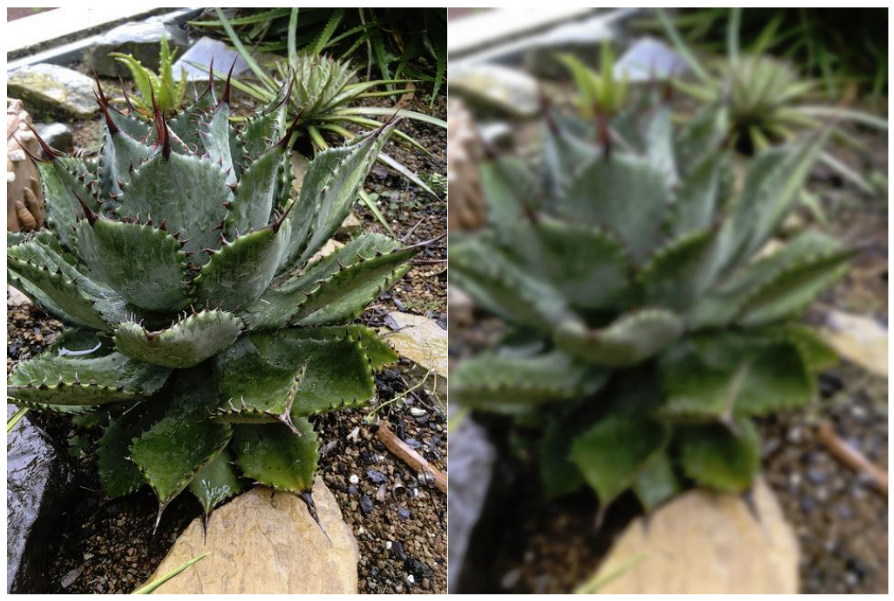}
	\end{tabular}
	\end{center}
   \caption[example] 
   { \label{fig:gaussian} 
Gaussian Blur of an Image with Sharp Edges.}
   \end{figure}

Figure \ref{fig:gaussian} demonstrates the effect of applying a Gaussian blur on an image with sharp edges. It shows more noticeable changes in high-frequency areas, such as in rocky dirt or the edges of the leaves than changes in the leaves. The needles on top of the leaves are high-frequency components because their colors and brightness greatly contrast from their background; therefore, they almost disappear after applying the Gaussian blur. 

In visual replay attacks, a streaming video out of synchronization with its audio could potentially raise suspicions to people monitoring the surveillance. To mask the video stream without be detected, the audio stream should also include a static noiseless recording. For example, pre-recorded video frames might have surrounding noises which are not in the current frames. 
When the video frames are replayed, a continuous audio stream replaying the same noise will be suspicious. Similarly, for an audio recording with no noise, the video should represent a static scene with no motion. The video frames and the audio samples which were collected independently and replayed together would represent a static scene with no background noise in the scene. The combination of both streams in this way would be realistic enough to deceive the security personnel monitoring the surveillance feed.

The duplicated audio samples are collected based on the noise detected in the audio of the live stream. Here, the audio is processed in parallel independent of the video frames. A Fast Fourier Transform (FFT) is performed on the samples to obtain a frequency domain representation of the input audio stream. Noise detection is performed by taking the mean volume across all frequencies and comparing it to a threshold. The threshold for audio is also decided based on different environmental settings of the camera. The FFT of a time domain signal $x(n)$ is,
\[ F(\omega) = \int_{-\infty}^{\infty} x(n) e^{-\omega x} dx \]

\noindent where \( \omega \) is the frequency of the Fourier transform.

For discrete signals, the Discrete Fourier Transform (DFT) \( A_k \) is used, for discrete samples \( a_n \) where $n = 0,\cdots, N-1$:

\[  A_k = \sum_{n=0}^{N-1} e^{-j(\frac{2\pi}{N})kn} a_n \]

\noindent where $N$ is the number of samples, and \( a_n \) is the sample value from the continuous signal \( x_n \). 



\subsection{Triggering Mechanisms}

In the literature, most of the previously tested frame duplication attacks used an editing software to manually alter the video content with the input of a database of previously recorded surveillance videos. With the increase in processing power of edge devices, the attacks can be tailored on site as required. An automated triggering mechanism is introduced here to implement the frame duplication attack in real time. In this paper, face detection (of the intruder) is employed as a trigger to launch the attack. Modern surveillance cameras produce high-resolution video streams, whose frames can be utilized to detect human faces. The frames per second (FPS) of the input high resolution video stream for face detection algorithm is lower, but a single frame with proper face detected is enough to trigger the attack, such that the FPS of video is irrelevant. 

As a part of our face recognition module, a histogram of oriented gradients (HOG) descriptor is used for fast human/face detection \cite{dalal2005histograms, zheng2018multispectral}. The gradient for human faces is trained using a machine learning (ML) algorithm, and each face has its unique encoding. The encoding vector is used later for detecting the appropriate face. In our attack, the face encoding vector of the intruder is generated in advance. We select a fast-human detection model, which uses a HOG cascade \cite{zhu2006fast}. A match with the stored face encoding vector (when the intruder shows up in the camera and is detected) is treated as a trigger of the replay attack. Moreover, a manual triggering mechanism is provided as an option to trigger the frame duplication attack. This allows more control over the deployment of the attack at the cost of connection to a attacker's remote server.

We have used face detection only as a demonstration of a triggering mechanism. For the given scenario, the trigger face is recorded in the videos and these recordings upon detecting a forgery can be used to identify the perpetrator. There are many other triggering methods, for example, a special QR code on a T-shirt, a unique gesture based trigger or even voice activated trigger. The main idea behind this approach is to demonstrate the remote capability of a trained algorithm to detect these secret triggers without communicating constantly with a rogue server. Post the trigger detection, different functionality can be carried out. In this paper, a frame duplication attack is performed automatically; other types of forgery include pixelating a person's face and shutting down a camera live feed. For the cameras with pan-tilt-zoom (PTZ) mechanisms the cameras can be pointed toward the opposite direction. 

\section{Experimental Study}
\label{sec:testbed}
\subsection{Testbed Setup}

A Raspberry Pi 3 board is used as an edge device, which represents modern surveillance cameras with equivalent computational power. For a higher resolution video stream, a Logitech C920 webcam is selected and a remote camera server is set up on the Raspberry Pi device and deployed in a network. The frame duplication attack algorithm is written in python scripting language. With the multi-threading support, the video replay attack code and audio replay attack code run in parallel to collect frames and samples. 

As shown in the earlier flow diagram of the algorithm in Fig. \ref{fig:state_diagram}, upon detecting a trigger, the \textit{video replay module} records the static background and maintains a collection of frame which reflects recent changes in the scene. It uses motion detection module to start and stop a static recording, and upon triggering the attack, the most recent recording is used to mask the live feed. 
Even if there is no object physically moving in the frame, changes in light intensity are also treated as motion; hence, switching on/off a light or an ambient light from a different room is considered as well. A video recording made during the day cannot be replayed at night for the attack, as it would otherwise easily expose the attacking system. 

The \textit{motion detection module} a part of  \textit{video replay module}, is responsible to differentiate between a static scene and the scene where objects or people are in motion. It uses Gaussian Blur on the frames to smooth out edges and minimize errors due to noise. The differences between subsequent frames are measured based on the changes in pixel values, which are compared with a threshold. 
For motion detection in indoor environments, the threshold on each pixel value change is set to 10. As a result, light brightness that changes very slowly is not considered as change, whereas turning on/off light is counted as a change. 
When over 0.5\% pixels have been changed, it is treated as motion detected. The threshold change is evaluated for indoor environments for optimal results. The Gaussian blur is implemented using the Python computer vision library, where the dimension of the Gaussian blur is $21 \times 21$, which determines the Gaussian kernel size.
   
The \textit{audio replay module} runs in parallel to \textit{video replay module}, records audio samples with no background noise and maintain a recent audio recording of the environment. It uses FFT to get the average value on the collected block of samples and compares it with other frequencies. The preset threshold determines if there is any noise detected in the background and starts a duplicated audio recording. With the trigger mechanism as \textit{face detection module}, when there is a face encoding match, the attack is triggered. The recent recordings of video and audio are used to mask the live feed when a static scene appears again. 


\subsection{Experimental Demonstration}

The frame duplication attack is tested in several scenarios, and the test results are illustrated in the following figures. Figure \ref{fig:face_Detection} shows the attack triggering mechanism. The algorithm detects and recognizes the intruder's face by matching it with a pre-generated encoding; then it waits for a static scene, until then it can mask the live feed with the duplicated feed. As discussed earlier, different triggering mechanisms can be used to trigger this attack and the resulting changes can be different based on the intentions of the intruder. This is a simple demonstration of trigger detection and attack effect. To hide the intruder's identity, real-time face blurring may also be applied (using a Gaussian blur module) on the bounding box around the detected face. However, an abrupt change in the video feed, e.g., blur around a person or a person suddenly disappearing could raise suspicions. 

 \begin{figure} [ht]
   \begin{center}
   \begin{tabular}{c} 
   \includegraphics[height=3.5cm]{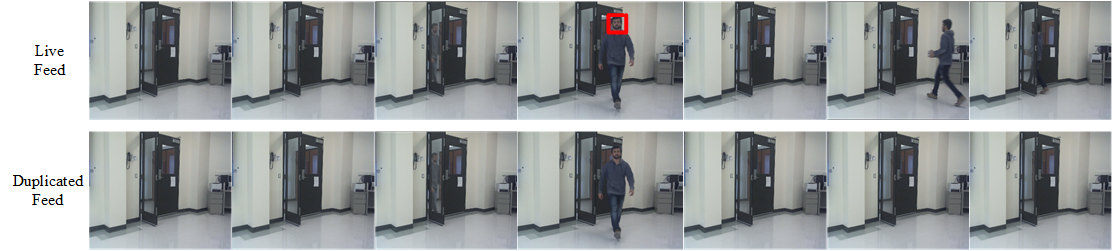}
	\end{tabular}
	\end{center}
   \caption[example] 
   { \label{fig:face_Detection} 
Face detection trigger and frame duplication attack deployed.}
   \end{figure}

    \begin{figure} [ht]
   \begin{center}
   \begin{tabular}{c} 
   \includegraphics[height=4.5cm]{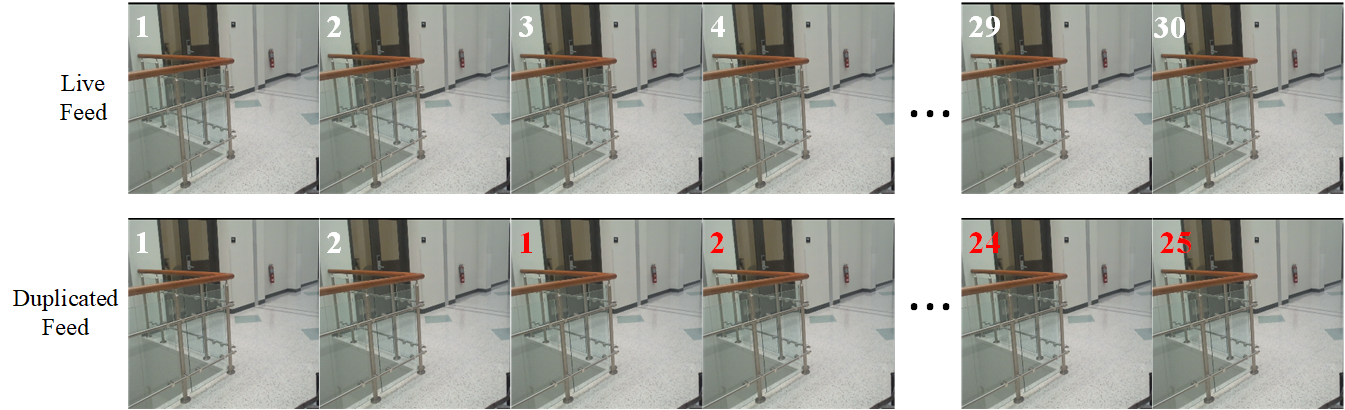}
	\end{tabular}
	\end{center}
   \caption[example] 
   { \label{fig:static_scene} 
Constantly record static scenes to accommodate the environmental changes and retain the latest frames.}
   \end{figure} 

With many possible changes in the environment, it would be easy to detect the duplicated frames if the changes were not carried over to the duplicated recording. To address this problem, the algorithm regularly checks for changes in the static background of the video feed. To keep up with the latest recording, the algorithm records a static scene whenever there is no motion detected and replaces it with the previous recording. Figure \ref{fig:static_scene} illustrates the frames in a static scene, and the attack was manually triggered. 
Here frames 1 and 2 are used as duplicated frames to mask frames 3 and 4. As the video stream proceeds, the algorithm updates previous duplicated frames 1 and 2 with new video frames 24 and 25 even if there were no changes in the environment. This constant update of duplicated frames is able to reflect new changes in the environment, which may or may not be perceivable to human eyes.

  \begin{figure} [t]
   \begin{center}
   \begin{tabular}{c} 
   \includegraphics[height=5.6cm]{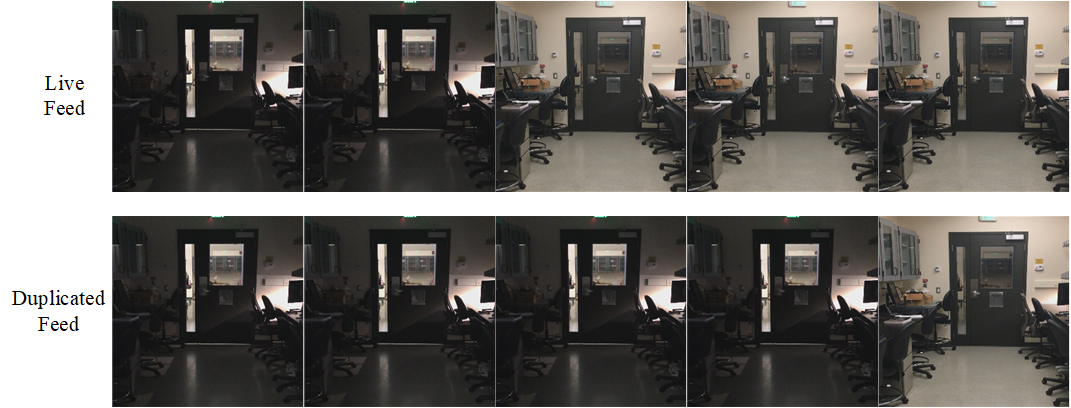}
	\end{tabular}
	\end{center}
   \caption[example] 
   { \label{fig:lights} 
Detecting changes in light intensity and covering up the change with duplicated frames.}
   \end{figure}

Light intensity changes are also considered as motion is detected. Once the light is on, however, the background is static again, and the new recording reflects the changes. 
Figure \ref{fig:lights} shows the scenarios of changes in lighting conditions.

\section{Real-Time Frame Duplication Attack Detection using ENF Signals}
\label{sec:enf}

The frame duplication algorithm is a real-time frame modification technique that masks the current events in a given scenario. Since the attack is application specific, frame duplication detection at a later stage from the attack event would be rendered counterproductive. Techniques which use camera sensor noise to identify the manufacturing source of the camera are not applicable since the duplicated video recordings were made by the same camera with all camera features set the same as in original recordings. A real-time watermarking protection algorithm could be a possible way to detect if there was any tampering with the watermark, but the attack is assumed in the scenario before the frames are sent to the server or database, i.e., the frames are duplicated before the watermark is applied. 

In order to detect these changes in frames collected in real time, a novel detection algorithm is proposed. Using electrical network frequency (ENF) fingerprints in the digital multimedia recordings, the method is capable of detecting the changes as soon as an attack is triggered. Due to the limited space, in this paper only the basic ideas are introduced, and interested readers are referred to another paper \cite{nagouthu2019detecting} for more comprehensive description of the algorithm, system implementation and experimental results.   

\subsection{Introduction to ENF Signals}

Electric Network Frequency (ENF) is a frequency fluctuation in the main power supply within each power grid \cite{kajstura2005application}. The nominal ENF is 60Hz in the United States and some parts of Japan, and 50Hz for most Asian and European countries. A typical ENF frequency is assumed to be at its nominal frequency, but due to the varying power supply and demand, there is a constant fluctuation caused by the generator trying to compensate for the requirements. These fluctuations in the electric frequency are instantaneous and called as an ENF signal. The ENF signal is observed to have similar fluctuations throughout an entire power grid. For our study purpose, we refer to the ENF signal in Eastern Interconnect of the United States. The fluctuations in ENF are within a certain range for different locations, and the US has fluctuation range of \( \pm 0.02\) Hz.

The fluctuations of ENF can be embedded in digital media recording due to the electromagnetic interference \cite{grigoras2005digital} or audible hum from many powered devices \cite{fechner2014humming, chai2013source}. The digital media recordings can either be an audio recording which has a high sampling rate and is capable of capturing ENF fluctuations in different harmonics, or video recordings where the source of ENF fluctuations come from varying light intensities running on a main power supply. The phenomenon of ENF signal embedded in different multimedia recordings has led to forensic analysis \cite{kajstura2005application, cooper2008electric} that uses a most commonly occurring and highly available ENF signal. It has been further extended in finding the geographical location \cite{garg2013geo} of a recording as well as timestamp verification of the recording \cite{garg2012modeling}.

\subsection{Proposal: an ENF Signal based Real-Time Detection}

The frequently occurring fluctuations in the ENF and the traces in the digital multimedia recordings are used as motivation for authenticating the surveillance recording online. Many frequency extraction techniques are used to extract the ENF signal in multimedia recordings like Short-time Fourier transform (STFT) \cite{grigoras2005digital}, Time-recursive iterative adaptive approach, 
rolling shutter and super-pixel based approaches in CMOS and CCD sensor cameras. 

Our proposed detection system consists of extracting the ENF signal from the audio-video recording at the edge device and using a standard reference database of ENF fluctuations collected using power recordings from the main supply. 
ENF fluctuations are inferred to be similar at different locations at the same time instant, and these ENF traces are embedded in media recordings through various factors. The ENF estimations from the power and the audio recordings are estimated simultaneously, and a correlation coefficient is used to evaluate the signal similarity. A low correlation coefficient indicates that the signals are not similar, which in turn implies the potential existence of maliciously injected duplicated frames. A sliding window based approach is proposed for online detection, and different parameter values are investigated to obtain the best setting. The experimental study has validated the effectiveness and correctness of the proposed technique\cite{nagouthu2019detecting}.

\section{Conclusions and Discussions}
\label{sec:conclusion}

The increased number of attacks on VSS may compromise the purpose of  installing public safety surveillance systems. The frame duplication attacks allow intruders to hide or escape from the monitor. The detection algorithms in the literature are not able to handle more complex attacks tailored for edge devices. The primary motivation behind this work is to address the security threats as soon as possible to minimize the foreseeable damage caused by the security breach.  

This paper demonstrated the feasibility of frame duplication attacks in the IoT applications, and proposed a novel real-time detection technique at the edge. Different surveillance scenarios were considered and the attacks were illustrated along with a triggering mechanism. The algorithm was demonstrated mainly in indoor environments and few outdoor scenarios. In case of a busy street with people and motor vehicles that are constantly moving, the proposed frame duplication technique might have difficulty to maintain a duplicated static record of the scenario. 
The proposed real-time detection scheme leverages the ENF signals as an environmental watermark. A preliminary concept proof prototype has been built and the testing results are encouraging \cite{nagouthu2019detecting}, which shows that an efficient and effective detection is feasible at the edge to deter the attackers before any serious consequences have been caused. 

 

\bibliography{report} 

\begin{thebibliography}{10}

\bibitem{collins2000system}
Collins, R.~T., Lipton, A.~J., Kanade, T., Fujiyoshi, H., Duggins, D., Tsin,
  Y., Tolliver, D., Enomoto, N., Hasegawa, O., Burt, P., et~al., ``A system for
  video surveillance and monitoring,'' {\em VSAM final report} ,  1--68 (2000).

\bibitem{nikouei2018real}
Nikouei, S.~Y., Chen, Y., Song, S., Xu, R., Choi, B.-Y., and Faughnan, T.~R.,
  ``Real-time human detection as an edge service enabled by a lightweight
  cnn,'' in [{\em 2018 IEEE International Conference on Edge Computing
  (EDGE)}{\nolinebreak\hspace{0.1em}]},   125--129, IEEE (2018).

\bibitem{yu2018pulic}
Yu, W., Xu, H., Nguyen, J., Blasch, E., Hematian, A., and Gao, W., ``Public
  safety communications: Survey of user-side and network-side solutions and
  future directions,'' {\em IEEE Access}~{\bf 6}(1),  70397--70425 (2018).

\bibitem{chen2016smart}
Chen, N., Chen, Y., Song, S., Huang, C.-T., and Ye, X., ``Smart urban
  surveillance using fog computing,'' in [{\em 2016 IEEE/ACM Symposium on Edge
  Computing (SEC)}{\nolinebreak\hspace{0.1em}]},   95--96, IEEE (2016).

\bibitem{nikouei2018smart}
Nikouei, S.~Y., Chen, Y., Song, S., Xu, R., Choi, B.-Y., and Faughnan, T.,
  ``Smart surveillance as an edge network service: From harr-cascade, svm to a
  lightweight cnn,'' in [{\em 2018 IEEE 4th International Conference on
  Collaboration and Internet Computing (CIC)}{\nolinebreak\hspace{0.1em}]},
  256--265, IEEE (2018).

\bibitem{costin2016security}
Costin, A., ``Security of cctv and video surveillance systems: Threats,
  vulnerabilities, attacks, and mitigations,'' in [{\em Proceedings of the 6th
  international workshop on trustworthy embedded
  devices}{\nolinebreak\hspace{0.1em}]},   45--54, ACM (2016).

\bibitem{mendez2017internet}
Mendez, D.~M., Papapanagiotou, I., and Yang, B., ``Internet of things: Survey
  on security and privacy,'' {\em arXiv preprint arXiv:1707.01879}  (2017).

\bibitem{nagothu2018microservice}
Nagothu, D., Xu, R., Nikouei, S.~Y., and Chen, Y., ``A microservice-enabled
  architecture for smart surveillance using blockchain technology,'' {\em arXiv
  preprint arXiv:1807.07487}  (2018).

\bibitem{xu2018real}
Xu, R., Nikouei, S.~Y., Chen, Y., Polunchenko, A., Song, S., Deng, C., and
  Faughnan, T.~R., ``Real-time human objects tracking for smart surveillance at
  the edge,'' in [{\em 2018 IEEE International Conference on Communications
  (ICC)}{\nolinebreak\hspace{0.1em}]},   1--6, IEEE (2018).

\bibitem{nikouei2018eiqis}
Nikouei, S.~Y., Chen, Y., Aved, A., and Blasch, E., ``Eiqis: Toward an
  event-oriented indexable and queryable intelligent surveillance system,''
  {\em arXiv preprint arXiv:1807.11329}  (2018).

\bibitem{obermaier2016analyzing}
Obermaier, J. and Hutle, M., ``Analyzing the security and privacy of
  cloud-based video surveillance systems,'' in [{\em Proceedings of the 2nd ACM
  International Workshop on IoT Privacy, Trust, and
  Security}{\nolinebreak\hspace{0.1em}]},   22--28, ACM (2016).

\bibitem{chen2018smart}
Chen, N. and Chen, Y., ``Smart city surveillance at the network edge in the era
  of iot: opportunities and challenges,'' in [{\em Smart
  Cities}{\nolinebreak\hspace{0.1em}]},   153--176, Springer (2018).

\bibitem{wu2015pseudo}
Wu, R., Liu, B., Chen, Y., Blasch, E., Ling, H., and Chen, G.,
  ``Pseudo-real-time wide area motion imagery (wami) processing for dynamic
  feature detection,'' in [{\em 2015 18th International Conference on
  Information Fusion (Fusion)}{\nolinebreak\hspace{0.1em}]},   1962--1969, IEEE
  (2015).

\bibitem{singh2015detection}
Singh, V.~K., Pant, P., and Tripathi, R.~C., ``Detection of frame duplication
  type of forgery in digital video using sub-block based features,'' in [{\em
  International Conference on Digital Forensics and Cyber
  Crime}{\nolinebreak\hspace{0.1em}]},   29--38, Springer (2015).

\bibitem{ulutas2017frame}
Ulutas, G., Ustubioglu, B., Ulutas, M., and Nabiyev, V., ``Frame
  duplication/mirroring detection method with binary features,'' {\em IET Image
  Processing}~{\bf 11}(5),  333--342 (2017).

\bibitem{bouchrika2018survey}
Bouchrika, I., ``A survey of using biometrics for smart visual surveillance:
  Gait recognition,'' in [{\em Surveillance in
  Action}{\nolinebreak\hspace{0.1em}]},   3--23, Springer (2018).

\bibitem{matusek2008nivss}
Matusek, F., Sutor, S., Kraus, K., Kruse, F., and Reda, R., ``Nivss: a nearly
  indestructible video surveillance system,'' in [{\em 2008 The Third
  International Conference on Internet Monitoring and
  Protection}{\nolinebreak\hspace{0.1em}]},   98--102, IEEE (2008).

\bibitem{wang2007exposing}
Wang, W. and Farid, H., ``Exposing digital forgeries in video by detecting
  duplication,'' in [{\em Proceedings of the 9th workshop on Multimedia \&
  security}{\nolinebreak\hspace{0.1em}]},   35--42, ACM (2007).

\bibitem{wahab2014passive}
Wahab, A. W.~A., Bagiwa, M.~A., Idris, M. Y.~I., Khan, S., Razak, Z., and
  Ariffin, M. R.~K., ``Passive video forgery detection techniques: a survey,''
  in [{\em 2014 10th International Conference on Information Assurance and
  Security}{\nolinebreak\hspace{0.1em}]},   29--34, IEEE (2014).

\bibitem{fadl2018authentication}
Fadl, S.~M., Han, Q., and Li, Q., ``Authentication of surveillance videos:
  detecting frame duplication based on residual frame,'' {\em Journal of
  forensic sciences}~{\bf 63}(4),  1099--1109 (2018).

\bibitem{heffner2013exploiting}
Heffner, C. and Solutions, T.~N., ``Exploiting surveillance cameras,'' {\em
  Tactical Network Solutions, Tech. Rep.}  (2013).

\bibitem{shoshitaishvili2015firmalice}
Shoshitaishvili, Y., Wang, R., Hauser, C., Kruegel, C., and Vigna, G.,
  ``Firmalice-automatic detection of authentication bypass vulnerabilities in
  binary firmware.,'' in [{\em NDSS}{\nolinebreak\hspace{0.1em}]},  (2015).

\bibitem{seralathan2018iot}
Seralathan, Y., Oh, T.~T., Jadhav, S., Myers, J., Jeong, J.~P., Kim, Y.~H., and
  Kim, J.~N., ``Iot security vulnerability: a case study of a web camera,'' in
  [{\em 2018 20th International Conference on Advanced Communication Technology
  (ICACT)}{\nolinebreak\hspace{0.1em}]},   172--177, IEEE (2018).

\bibitem{xie2017vulnerability}
Xie, W., Jiang, Y., Tang, Y., Ding, N., and Gao, Y., ``Vulnerability detection
  in iot firmware: A survey,'' in [{\em 2017 IEEE 23rd International Conference
  on Parallel and Distributed Systems (ICPADS)}{\nolinebreak\hspace{0.1em}]},
  769--772, IEEE (2017).

\bibitem{dalal2005histograms}
Dalal, N. and Triggs, B., ``Histograms of oriented gradients for human
  detection,'' in [{\em international Conference on computer vision \& Pattern
  Recognition (CVPR'05)}{\nolinebreak\hspace{0.1em}]},   {\bf 1},  886--893,
  IEEE Computer Society (2005).

\bibitem{zheng2018multispectral}
Zheng, Y., Blasch, E., and Liu, Z.,  [{\em Multispectral Image Fusion and
  Colorization}{\nolinebreak\hspace{0.1em}]}, SPIE Press (2018).

\bibitem{zhu2006fast}
Zhu, Q., Yeh, M.-C., Cheng, K.-T., and Avidan, S., ``Fast human detection using
  a cascade of histograms of oriented gradients,'' in [{\em 2006 IEEE Computer
  Society Conference on Computer Vision and Pattern Recognition
  (CVPR'06)}{\nolinebreak\hspace{0.1em}]},   {\bf 2},  1491--1498, IEEE (2006).

\bibitem{nagouthu2019detecting}
Nagothu, D., Chen, Y., Blasch, E., Aved, A., and Zhu, S., ``Detecting malicious
  false frame injection attacks on the internet of video things using
  electrical network frequency signals,'' {\em Sensors, Special Issue on
  Intelligent Signal Processing, Data Science and the IoT World} ,  1--20
  (2019).

\bibitem{kajstura2005application}
Kajstura, M., Trawinska, A., and Hebenstreit, J., ``Application of the
  electrical network frequency (enf) criterion: A case of a digital
  recording,'' {\em Forensic science international}~{\bf 155}(2-3),  165--171
  (2005).

\bibitem{grigoras2005digital}
Grigoras, C., ``Digital audio recording analysis--the electric network
  frequency criterion,'' {\em International Journal of Speech Language and the
  Law}~{\bf 12}(1),  63--76 (2005).

\bibitem{fechner2014humming}
Fechner, N. and Kirchner, M., ``The humming hum: Background noise as a carrier
  of enf artifacts in mobile device audio recordings,'' in [{\em IT Security
  Incident Management \& IT Forensics (IMF), 2014 Eighth International
  Conference on}{\nolinebreak\hspace{0.1em}]},   3--13, IEEE (2014).

\bibitem{chai2013source}
Chai, J., Liu, F., Yuan, Z., Conners, R.~W., and Liu, Y., ``Source of enf in
  battery-powered digital recordings,'' in [{\em Audio Engineering Society
  Convention 135}{\nolinebreak\hspace{0.1em}]},  Audio Engineering Society
  (2013).

\bibitem{cooper2008electric}
Cooper, A.~J., ``The electric network frequency (enf) as an aid to
  authenticating forensic digital audio recordings--an automated approach,'' in
  [{\em Audio Engineering Society Conference: 33rd International Conference:
  Audio Forensics-Theory and Practice}{\nolinebreak\hspace{0.1em}]},  Audio
  Engineering Society (2008).

\bibitem{garg2013geo}
Garg, R., Hajj-Ahmad, A., and Wu, M., ``Geo-location estimation from electrical
  network frequency signals,'' in [{\em 2013 IEEE International Conference on
  Acoustics, Speech and Signal Processing}{\nolinebreak\hspace{0.1em}]},
  2862--2866, IEEE (2013).

\bibitem{garg2012modeling}
Garg, R., Varna, A.~L., and Wu, M., ``Modeling and analysis of electric network
  frequency signal for timestamp verification,'' in [{\em 2012 IEEE
  International Workshop on Information Forensics and Security
  (WIFS)}{\nolinebreak\hspace{0.1em}]},   67--72, IEEE (2012).

\end{thebibliography}
\bibliographystyle{spiebib} 

\end{document}